\documentstyle[11pt,psfig]{article}
\begin{document}

\begin{center}
{\Large\bf Kaon photo-production on the nucleon and deuteron}\vspace{6mm}\\
P. BYD\v{Z}OVSK\'Y\footnote{E-mail: bydz@ujf.cas.cz}, M. SOTONA\\
{\it Nuclear Physics Institute, Czech Academy of Sci.\\
25068 \v{R}e\v{z} near Prague, Czech Republic}\vspace{3mm}\\
O. HASHIMOTO, T. TAKAHASHI\\
{\it Department of Physics, Tohoku University\\
Sendai, 980-8578, Japan}   
\end{center}

\begin{abstract}Isobaric models for the photo-production of K$^+$ are 
discussed and their predictions are shown in the K$^0$ photo-production. 
The models are further used in spectator model calculations of 
the K$^0$ photo-production on deuteron. Considerable dependence of 
the inclusive cross section on the elementary amplitude was found.
\end{abstract} 

\section{Introduction}

The photo-production of kaons on nucleons has been studied intensively 
last years [1-10]. Analysis of the process 
contributes to our understanding of dynamics in the strange sector. 
Special attention is payed to investigation of a resonance content of 
the amplitudes, particularly searching for ``missing resonances'', 
the structure of hadrons (form factors), and to better fixing 
the effective couplings. This is possible especially after copious 
and good quality data were collected in JLab (CLAS) \cite{CLAS}, 
ELSA (SAPHIR) \cite{SAPHIR03}, and SPring-8 (LEPS) \cite{LEPS}. 
The amplitude of the process is also an input information in calculations 
of excited spectra in the hypernucleus photo-production \cite{Mot03}. 
A good quality description of the elementary process can then minimise 
a theoretical uncertainty of the hypernuclear results. 

There are several approaches to treat the elementary process. Among them 
the isobaric models based on the effective description utilizing only the 
hadronic degrees of freedom are suitable for their further use in more 
complex calculations. Other approaches are eligible either for higher 
energies ($E_{\gamma}>4$ GeV), the Regge model \cite{Regge}, or to the 
threshold region, the Chiral Perturbation Theory \cite{ChPT}. 
Quark models \cite{SagQM} are too complicated for their further use in 
the hypernuclear calculations. 

While there are many models which provide a satisfactory description 
of data in the p($\gamma$,K$^+$)$\Lambda$ reaction, almost nothing in 
known about the K$^0$ photo-production, the only few data being in 
the p($\gamma$,K$^0$)$\Sigma^+$ channel \cite{Goe99}. However, the first 
measurements of the K$^0$ photo-production from carbon and deuteron in 
the threshold region were performed at Tohoku University \cite{Tsu04}. 
Utilizing these data the simple well known deuteron structure allows 
one to obtain information on the elementary process, 
n($\gamma$,K$^0$)$\Lambda$, which is difficult to obtain otherwise.  

\section{Photo-Production on the Nucleon}\label{prdN} 
 
In the effective hadronic Lagrangian approach various channels 
connected via the final state interaction have to be treated 
simultaneously to take unitarity properly into account \cite{Penner,Ch01}.  
In Ref.~\cite{Ch01} the coupled-channel approach has been used 
to include effects of $\pi$N intermediate states in 
the p($\gamma$,K$^+$)$\Lambda$ process. However, tremendous 
simplifications originate in neglecting the rescattering effects 
in the formalism assuming that they are included to some extent 
through effective values of the strong coupling constants fitted 
to data. This simplifying assumption was adopted in many of the 
isobaric models, e.g. Adelseck-Saghai (AS1) \cite{AS90}, 
William-Ji-Cotanch (WJC) \cite{WJC92}, Saclay-Lyon (SLA) \cite{SLA},
Kaon-MAID (K-MAID) \cite{Ben99}, and Janssen {\it et al.} \cite{Jan01}.

In these isobaric models the amplitude obtains contributions from 
the Born terms and exchanges of resonances. Due to absence of 
a dominant resonance in the process large number of possible 
combinations of the resonances with mass smaller than 2 GeV must 
be taken into consideration \cite{AS90,SLA}. This number of models 
is limited assuming constraints set by SU(3) symmetry \cite{AS90,SLA}, 
crossing symmetry \cite{WJC92,SLA}, and duality hypothesis \cite{WJC92}. 
Adopting the SU(3) constraints to the two main coupling constants, 
however, makes the contribution of the Born terms nonphysically 
large \cite{Jan01}. To reduce this contribution either hyperon 
resonances \cite{SLA} or hadronic form factors \cite{Ben99} must 
be added, or a combination of both \cite{Jan01}. The hadronic 
form factors which mimic a structure in the strong vertex are 
included in the K-MAID and Janssen models maintaining the gauge 
invariance of the amplitude \cite{DW01}. In the analysis we use 
also our models M2 and H2 presented in Ref.~\cite{HYP03}. In these 
models the SU(3) symmetry is assumed and the hadronic form factors 
were taken into account by the recipe of Ref.~\cite{DW01}. The models 
differ in the resonance content \cite{HYP03}. Free parameters 
were fitted to the ample set of CLAS data \cite{CLAS}. 
 
The strong coupling constants in the K$^0\Lambda$ and K$^+\Lambda$ 
channels are related via the SU(2) isospin symmetry: 
$g_{{\rm K}^+\Lambda {\rm p}}= g_{{\rm K}^0\Lambda {\rm n}}$ 
and $g_{{\rm K}^+\Sigma^0 {\rm p}}= -g_{{\rm K}^0\Sigma^0 {\rm n}}$. 
In the electromagnetic vertexes a ratio of the neutral to charged 
coupling constants have to be known. For the nucleon resonances it can 
be related to the known helicity amplitudes of the nucleons \cite{PDP} 
whereas in the $t$ channel it relates to the decay widths, which 
were measured only for the K$^*$ meson \cite{PDP}: 
r$_{\rm KK^*}=-\sqrt{\Gamma_{{\rm K}^{*0}\rightarrow {\rm K}^0\gamma}/ 
\Gamma_{{\rm K}^{*+}\rightarrow {\rm K}^+\gamma}}=-1.53$ where the sign 
was set from the quark model prediction. Since the decay widths of 
the K$_1$ meson are not known the appropriate ratio,  r$_{\rm KK_1}$, 
have to be fixed in the models. It was fitted to the K$^0\Sigma^+$ 
data in K-MAID \cite{Ben99}, r$_{\rm KK_1}=-0.45$, but in the other 
models it is {\it a free parameter}. The $u$ channel contributions 
remain unchanged in the K$^0$ photo-production. 

\section{Photo-Production on the Deuteron}

Our aim is to demonstrate a dependence of the inclusive cross section in 
d($\gamma$,K$^0$)$\Lambda$p process on the input elementary amplitudes. 
We show it in a simple model based on the impulse approximation 
in which the proton acts as a spectator. Since a part of the K$\Lambda$ 
interaction in the final state (FSI) is absorbed in the coupling 
constants of the elementary amplitude and the KN interaction is weak 
on the hadronic scale the main lack of precision comes from ignoring 
the $\Lambda$N FSI. However, it has been shown in Ref.~\cite{LiWr91} 
that the $\Lambda$N interaction is important for the exclusive process 
d($\gamma$,K$^+\Lambda$)n near the threshold but is not so significant 
for the inclusive d($\gamma$,K$^+$)$\Lambda$n one. Assuming that 
the nature of the FSI in the K$^+$ and K$^0$ production is not too much 
different we suppose our simple model as a good approach here.

In {\it the spectator approximation} we can write the cross section as 
\begin{eqnarray}
d^9\sigma = \frac{m_{\Lambda}m_{N'}}
{64\pi^4 P_{\gamma}\cdot P_d\,E_K E_{\Lambda} E_{N'}}
\int d^4P_N \;\delta^4(P^e_f-P^e_i)\, 
\frac{(s-m_N^2)^2}{m_{\Lambda}m_N}\;\times\nonumber\\
\times\;\frac{d\sigma^e}{dt}\,\delta^4(P_d-P_N-P_{N'})\; 
\frac{\frac{1}{6}\sum |M_{fi}|^2} {\frac{1}{4}\sum |M^e_{fi}|^2}
\;d^3p_K\,d^3p_{\Lambda}\,d^3p_{N'}\;,
\label{eqv3}
\end{eqnarray}
where we have added the integration over the four-momentum 
of the target nucleon $P_N$, splitting the $\delta$-function into 
two parts, and have introduced the invariant cross section for 
the two-body process 
\begin{equation}
\frac{d\sigma^e}{dt}(s,t) = \frac{1}{4\pi}\frac{m_Nm_{\Lambda}}
{(s-m_N^2)^2}\ \frac{1}{4}\ \sum_{} |M^e_{fi}|^2\; ,
\label{ICS}
\end{equation}
with $P^e_i = P_{\gamma} + P_N$, $P^e_f = P_K + P_{\Lambda}$, 
$s = (P_{\gamma} + P_N)^2$, and $t = (P_{\gamma} - P_K)^2$.  
The spin averaged matrix element $\frac{1}{6}\sum 
|M_{fi}|^2$ in (\ref{eqv3})  can be expressed via the elementary 
one in the deuteron laboratory frame 
\begin{equation}
\frac{1}{6} \sum_{}\;|M_{fi}|^2 = 
(2\pi)^3 \,\frac{2m_d E_{N'}}{E_N} \,\frac{1}{4} \sum_{} 
\,|M_{fi}^e|^2 \,u_d(p_{N'})^2\;,
\label{final}
\end{equation}
where we have followed the formalism of Ref.~\cite{Dar03}. 
We have assumed the isospin formalism and have added the isospin 
factor $\sqrt{2}$ to the elementary amplitude to take account of 
the antisymmetrization of two nucleons in the intermediate state.
In Eq.~(\ref{final}) the energy of the target nucleon is given 
by $E_N = E_d- E_{N'} = E_K + E_{\Lambda} - E_{\gamma}$, for the 
{\it off-shell} approximation and by $E_N = \sqrt{m_N^2 + \vec{p}_N^2}$
for the {\it on-shell} one. In the on-shell approximation the  
energy conservation law is obviously violated in the two-body 
vertex, $E_{\gamma} + E_N >  E_K + E_{\Lambda}$, as well as in 
the deuteron one, $E_N + E_{N'} > E_d$.

The final expression for the inclusive cross section in the lab 
frame is
\begin{equation}
\frac{d^3\sigma}{d|\vec{p}_K|\,d\Omega_K} =  
\int \!\!\frac{m_{N'}(s-m_N^2)^2\, \vec{p}_K^{\;2}\:|\vec{p}_{N'}|}
{4m_N E_{\gamma} E_K E_N |\vec{p}_{\gamma}-\vec{p}_K|} 
\frac{u_d(p_{N'})^2}{\pi}
\frac{d\sigma^e}{dt}\,d|\vec{p}_{N'}|\,d\Phi_{N'}'\;,
\label{crs}
\end{equation}
where all energies and momenta are in the lab frame, the target 
nucleon mass is taken to be consistent with kinematics, 
$m_N = \sqrt{P_N^2}$, $\vec{p}_N = -\vec{p}_{N'}$. In our 
calculations we used the non relativistic Bonn wave 
functions\cite{Bonn}.

\section{Results and Discussion}
%
%
\begin{figure}[ht]
\centerline{\psfig{figure=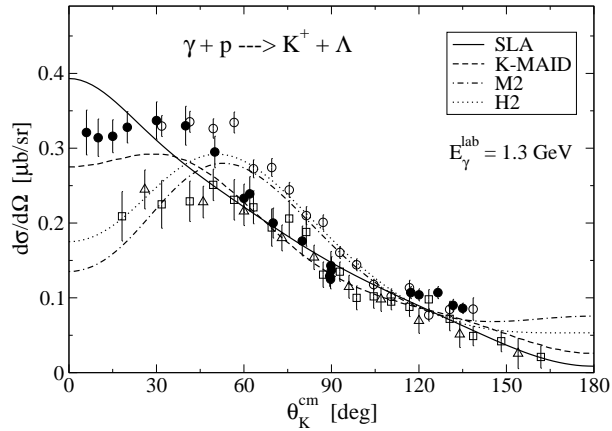,width=8.cm,angle=-90}}   
\caption{Calculated differential cross section in the photo-production 
of K$^+$ on proton are compared with data from 
Refs. \protect\cite{CLAS}(circle), \protect\cite{SAPHIR03}(square), 
\protect\cite{SAPHIR98}(triangle), and \protect\cite{BleckDec}(dot).
\label{figure1}}
\end{figure}
\begin{figure}[ht]
\centerline{\psfig{figure=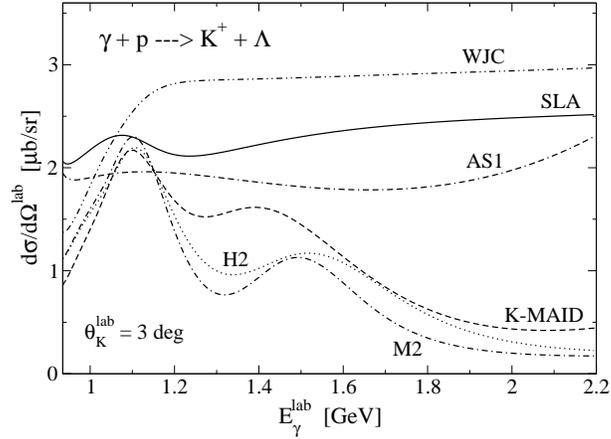,width=8.cm,angle=-90}}
\caption{Calculated differential cross sections for the forward 
angle are shown as a function of energy in the laboratory frame. 
\label{figure2}}
\end{figure}

The isobaric models being fitted to the data on the K$^+$ photo-production 
on proton are in satisfactory agreement with the data, except for  
the forward kaon angles, as it is demonstrated in Fig.~\ref{figure1} for 
the SLA, K-MAID, M2, and H2 models. The SLA and K-MAID models were 
fitted to the older data (triangles ans dots) whereas the M2 and H2 
ones were adjusted only to the latest CLAS data \cite{CLAS}. The obvious 
and most serious, in the view of the hypernuclear calculations, 
discrepancy of the results is at $\theta_{\rm K}^{\rm cm}<40$ deg. 
In this region, however, one observes an inconsistency of the 
experimental data too, further stressed by the new CLAS and SAPHIR data.  
The systematic discrepancy of results for a very small kaon angle 
is shown in Fig.~\ref{figure2}. Models with hadronic form factors, K-MAID, 
M2, and H2, provide much smaller cross sections at 
$E_{\gamma}^{\rm lab}>1.5$ GeV than the others.
%
%

%
%
\begin{figure}[ht]
\centerline{\psfig{figure=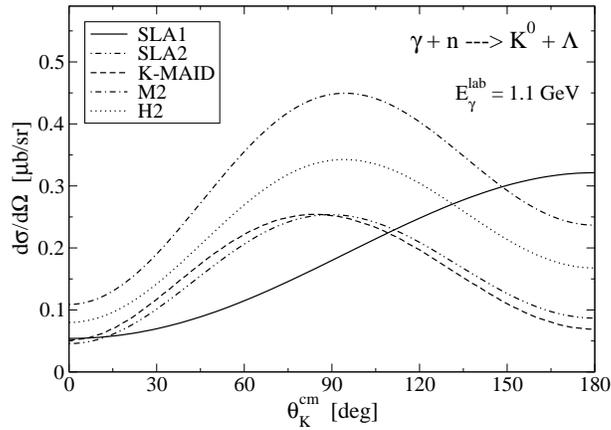,width=8.cm,angle=-90}}   
\caption{Predictions of the models for the differential cross section 
in the photo-production of K$^0$ on neutron are shown for the photon 
energy of 1.1 GeV.\label{figure3}}
\end{figure}
In Figure~\ref{figure3} predictions of the models for the K$^0$ 
photo-production on neutron are shown. As mentioned already in 
Sect.~\ref{prdN}, the only free parameter in 
this channel is the ratio r$_{\rm KK1}$. 
We have used the value of -0.45, as it was fixed in the 
K-MAID \cite{Ben99}, in the models M2 and H2 too. In the case of 
the SLA model we have found two values of the ratio, -1.6 (SLA1) and 
-3.4 (SLA2), which provide results very near to those of the K-MAID 
at forward angles, see Fig.~\ref{figure3}. The models reveal more 
different results than in the K$^+$ production. The bump structure 
is produced by the $N^*(1720)$ exchange.

%
%
\begin{figure}[ht]
\centerline{\psfig{figure=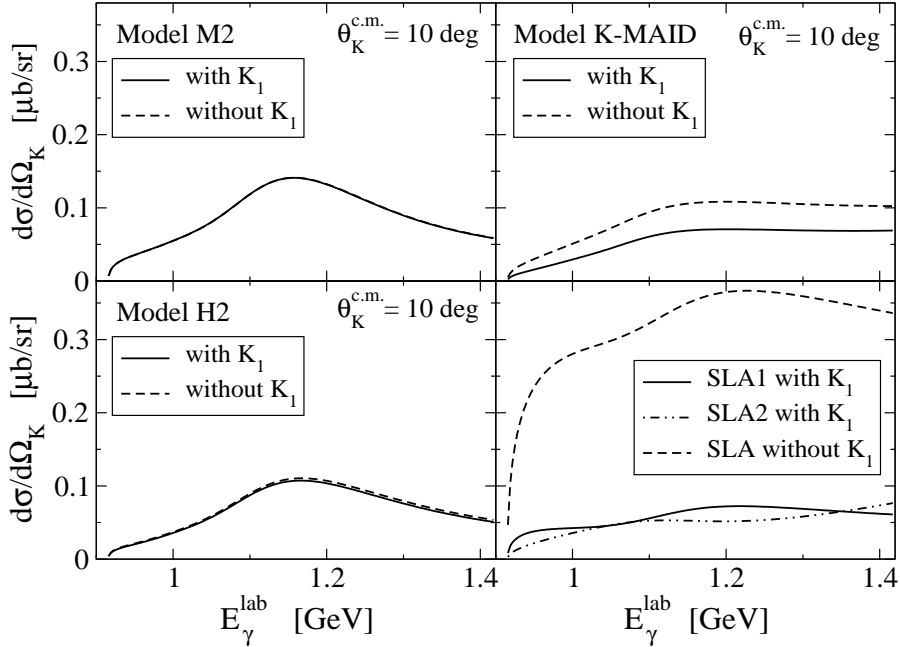,width=12.cm,angle=-90}}   
\caption{Contribution of the K$_1$ exchange to the differential cross 
section in the photo-production of K$^0$ on neutron is shown 
as a function of energy at kaon angle of 10 deg. 
\label{figure4} }
\end{figure}
In Figure~\ref{figure4} we demonstrate that the K-MAID and SLA are 
much more sensitive to a contribution of the K$_1$ exchange than 
the M2 and H2 models. This phenomenon is valid for energies up to 
1.5 GeV and kaon lab angles up to 50 deg. 
That makes a choice of values for r$_{\rm KK1}$ less important 
in the M2 and H2 than in the SLA and K-MAID models.

%
%
\begin{figure}[ht]
\centerline{\psfig{figure=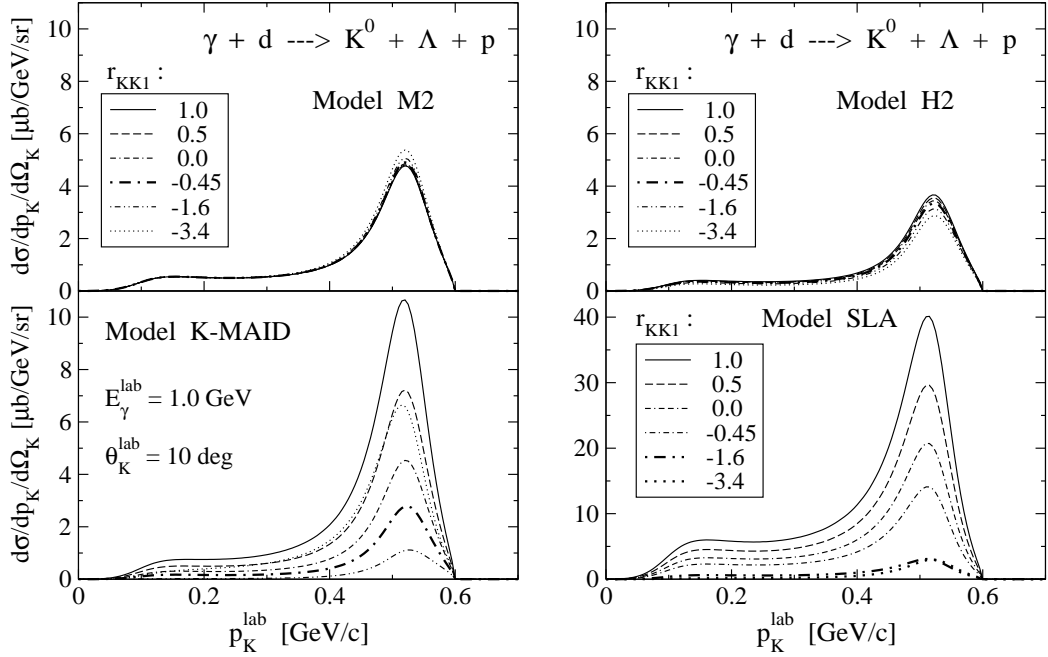,width=14.cm,angle=-90}}   
\caption{Double differential cross sections as a function of the kaon 
lab momentum for the photo-production of K$^0$ on deuteron are plotted 
in dependence of the ratio of electromagnetic couplings for K$_1$, 
r$_{\rm KK1}$, for the models M2, H2, K-MAID, and SLA. The fat lines   
indicate results for the parameters (see also the text) used 
in Fig~\protect\ref{figure6}. Calculations were performed at the photon 
lab energy of 1 GeV and kaon lab angle of 10 deg.\label{figure5} }
\end{figure}
A sensitivity of the cross section (\ref{crs}) to the r$_{\rm KK1}$ 
parameter is shown again for the d($\gamma$,K$^0$)$\Lambda$p reaction 
in Fig.~\ref{figure5}. The figure shows that the value of 
r$_{\rm KK1}$ is not too important for the M2 and H2, so that they 
provide really predictions in the K$^0$ channel at the kinematical 
region assumed here. On the contrary, results of the SLA model vary 
very strongly with values of r$_{\rm KK1}$ (notice the scale of 
the appropriate figure).   
%
%
\begin{figure}[ht]
\centerline{\psfig{figure=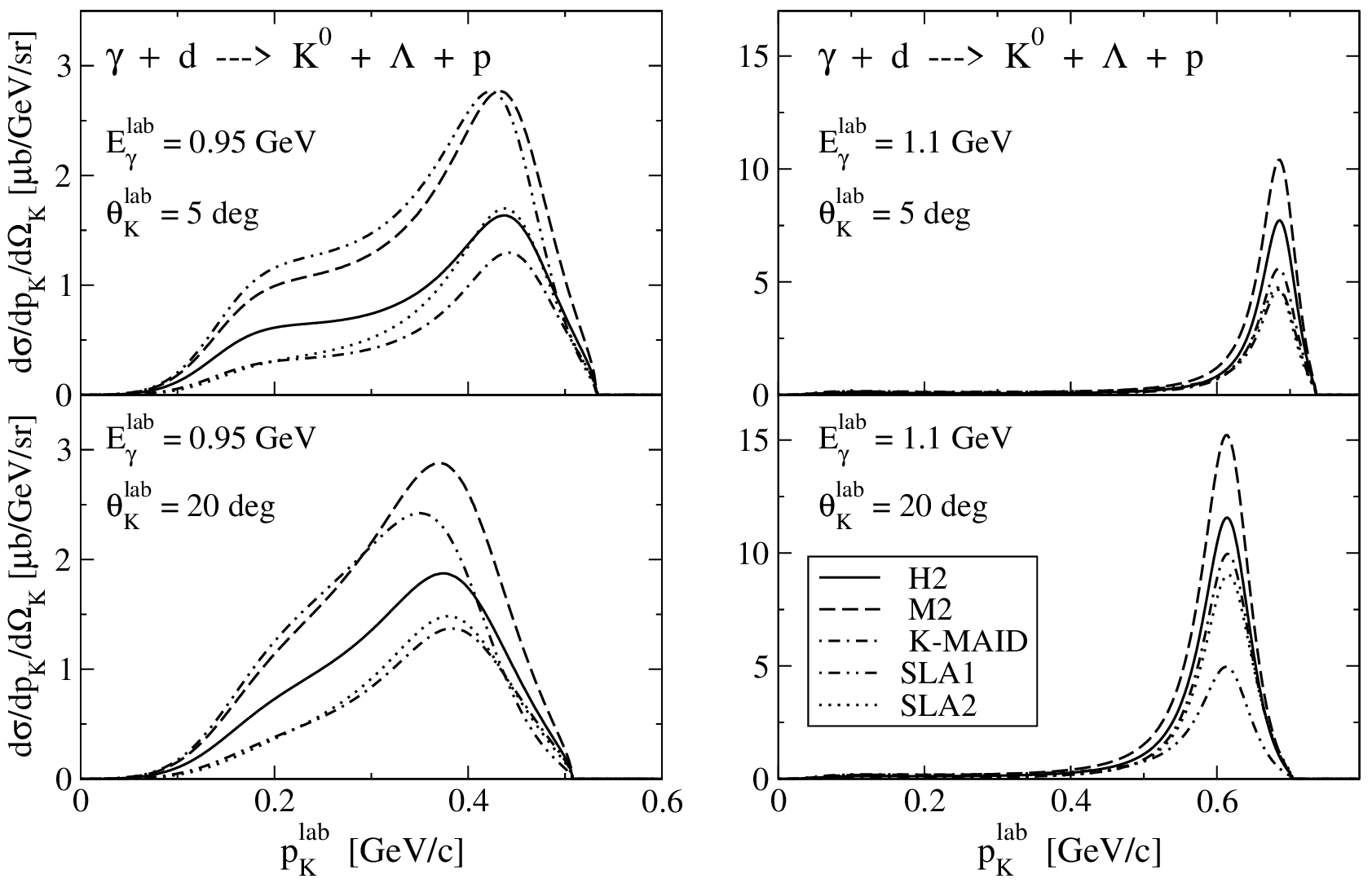,width=14.cm,angle=-90}}   
\caption{Calculated double differential cross sections in 
the photo-production of K$^0$ on deuteron are shown for two 
energies and angles. \label{figure6} }
\end{figure}
Predictions of the models for the inclusive cross section (\ref{crs}) 
are shown for the K$^0$ photo-production on the deuteron for 
small kaon lab angles and photon energies in Fig.~\ref{figure6}. 
The results differ significantly in some cases which enable the 
data from Tohoku experiment \cite{Tsu04} to discriminate between 
the elementary models.

To summarise, we showed that the isobaric models still provide different 
predictions for the cross section of the K$^+$ photo-production 
at forward angles which then causes large input uncertainty of 
the hypernuclear calculations. 
Predictions for the inclusive cross sections of the simple model for  
the d($\gamma$,K$^0$)$\Lambda$p reaction display considerable 
sensitivity to elementary amplitudes and therefore we expect 
that the experimental data will allow to discriminate between various 
elementary amplitudes which otherwise fit the p($\gamma$,K$^+$)$\Lambda$ 
data equally well.
\vspace{4mm}

\noindent {\small One of the authors (P.B.) wishes to thank the organisers 
for their kind invitation to this highly stimulating workshop. 
We are also very grateful to K. Miyagawa and A. Salam for very 
helpful discussions. This work was supported by grant 202/02/0930 of 
the Grant Agency of the Czech Republic.}

\end{document}